\documentclass[sigconf]{acmart}
\AtBeginDocument{%
  }

\setcopyright{none}
\acmYear{2026}
\acmDOI{}
\acmConference[LOCO2026]{2$^{nd}$ workshop on low carbon computing}{September 10--11,
  2026}{Lancaster, UK}
\acmISBN{}

\usepackage{subcaption}

\begin{document}

%%
%% The "title" command has an optional parameter,
%% allowing the author to define a "short title" to be used in page headers.
\title{Too cheap to matter: over abundant microchips, and what we can learn from them}

%%
%% The "author" command and its associated commands are used to define
%% the authors and their affiliations.
%% Of note is the shared affiliation of the first two authors, and the
%% "authornote" and "authornotemark" commands
%% used to denote shared contribution to the research.
\author{Adrian Friday}
% \authornote{Both authors contributed equally to this research.}
\email{a.friday@lancaster.ac.uk}
\orcid{0000-0002-9228-7365}
\correspondingauthor
\authornotemark[1]
\affiliation{%
  \institution{Lancaster University}
  \city{Lancaster}
  %\state{Lancashire}
  \country{UK}
}

\author{Fieke Jansen}
% \email{fieke@criticalinfralab.net}
\orcid{0000-0002-1295-7788}
\affiliation{%
  \institution{Critical infrastructure lab at University of Amsterdam}
  \city{Amsterdam}
  \country{Netherlands}}

\author{Gauthier Roussilhe}
% \email{gauthier.roussilhe@hubblo.org}
\orcid{0000-0003-4236-9343}
\affiliation{%
  \institution{Hubblo}
  % \city{???}
  \country{France}
}

\author{Srinjoy Mitra}
% \email{srinjoy.mitra@ed.ac.uk}
\orcid{0000-0003-1505-2316}
\affiliation{%
 \institution{University of Edinburgh}
 \city{Edinburgh}
 % \state{Arunachal Pradesh}
 \country{UK}}
%%
%% By default, the full list of authors will be used in the page
%% headers. Often, this list is too long, and will overlap
%% other information printed in the page headers. This command allows
%% the author to define a more concise list
%% of authors' names for this purpose.
\renewcommand{\shortauthors}{Friday et al.}

%%
%% The abstract is a short summary of the work to be presented in the
%% article.
\begin{abstract}
Ultra-cheap microchips (<\$1) are so abundant they've become a `smart material' integrated and disposable in everyday things.  Hidden in our everyday products, we have entirely lost sight of them, yet they account for the vast majority of the >400 billion pieces sold each year. As new technology nodes are released, older ones (from as far back as the 1980s) continue to be produced. These microchips do not exist on their own; they are packaged into every possible item to bring `smartness', necessary or not; this simultaneously increases their obsolescence.  While the latest ICs power our data centres and AI revolution that draws our attention, what about technology so disposable that it has become entirely invisible?  We report on our workshop at ICT4S exploring these devices' true costs, and pose challenges to the LOCO community to push back on this system, and develop the skills necessary to create lasting technology and avoid further e-Waste.
\end{abstract}

%%
%% The code below is generated by the tool at http://dl.acm.org/ccs.cfm.
%% Please copy and paste the code instead of the example below.
%%
% \begin{CCSXML}
% <ccs2012>
%  <concept>
%   <concept_id>00000000.0000000.0000000</concept_id>
%   <concept_desc>Do Not Use This Code, Generate the Correct Terms for Your Paper</concept_desc>
%   <concept_significance>500</concept_significance>
%  </concept>
%  <concept>
%   <concept_id>00000000.00000000.00000000</concept_id>
%   <concept_desc>Do Not Use This Code, Generate the Correct Terms for Your Paper</concept_desc>
%   <concept_significance>300</concept_significance>
%  </concept>
%  <concept>
%   <concept_id>00000000.00000000.00000000</concept_id>
%   <concept_desc>Do Not Use This Code, Generate the Correct Terms for Your Paper</concept_desc>
%   <concept_significance>100</concept_significance>
%  </concept>
%  <concept>
%   <concept_id>00000000.00000000.00000000</concept_id>
%   <concept_desc>Do Not Use This Code, Generate the Correct Terms for Your Paper</concept_desc>
%   <concept_significance>100</concept_significance>
%  </concept>
% </ccs2012>
% \end{CCSXML}

% \ccsdesc[500]{Do Not Use This Code~Generate the Correct Terms for Your Paper}
% \ccsdesc[300]{Do Not Use This Code~Generate the Correct Terms for Your Paper}
% \ccsdesc{Do Not Use This Code~Generate the Correct Terms for Your Paper}
% \ccsdesc[100]{Do Not Use This Code~Generate the Correct Terms for Your Paper}

%% A "teaser" image appears between the author and affiliation
%% information and the body of the document, and typically spans the
%% page.
\begin{teaserfigure}
 \includegraphics[width=\textwidth]{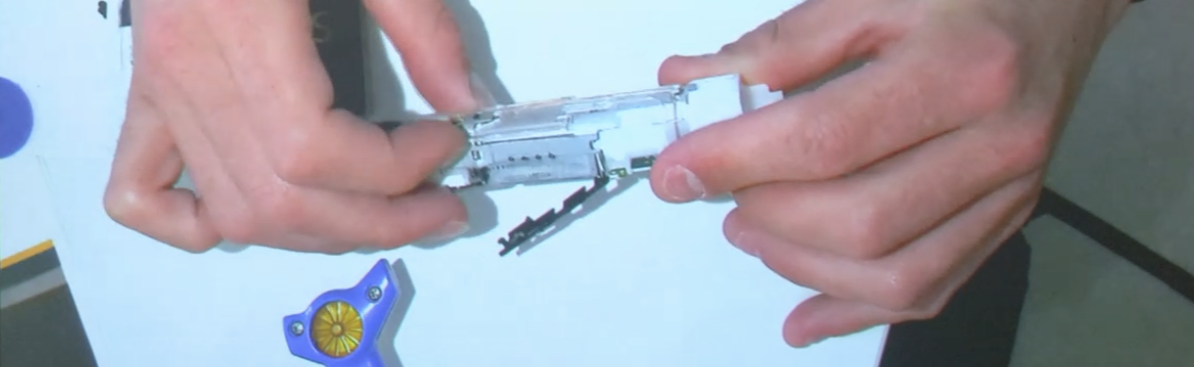}
  \caption{Embedded electronics and integrated circuits in a largely disposable plastic toy.}
  \label{fig:teaser}
\end{teaserfigure}

\received{30 June 2026}
% \received[revised]{12 March 2009}
% \received[accepted]{5 June 2009}

%%
%% This command processes the author and affiliation and title
%% information and builds the first part of the formatted document.
\maketitle

\section{Introduction}
Ultra-cheap microchips are very low-cost semiconductor integrated circuits (IC) so cheap and abundant they enable convenience functions in a multitude of everyday items around us. In this paper we set out to highlight the economic and environmental impact of these invisible but indispensable artefacts of modern life. Low cost ICs create the vast bulk of the >400 billion ICs sold each year. While the latest Nvidia GPU is now so sought after prices have sky rocketed such that they are nearly impossible to buy on the open market, the average cost at the low-end of the IC market is <\$1, and most are readily available today~\cite{StatistaMarketInsights}. Yet, this abundance comes with a huge ecological cost.

The average size of a microchip is around 2mm$^2$, and consists of a miniscule amount of silicon and other rare and exotic materials. Manufactured at such volume, these tiny inexpensive microchips are packaged into every possible gadget around us, whether necessary or not: blinking LEDs, sensing in toasters or allowing toys to talk.  Yet, their low cost and invisibility within other products increases the obsolescence of those items. These microchips are unrepairable by design, with even the circuit boards they are packaged into very often economically unviable to repair. From disposable vapes to smart toys, cheap microchips enable the production of billions of new objects that will be discarded after often very short life-cycles. For example, over one billion `fast-tech' gadgets (e.g., handheld fans) were bought by UK consumers alone in 2024, half of which have already been disposed of~\cite{banfield-nwachi2025its-cheap}. Some, like the one-use digital pregnancy test, are even single-use \emph{by design}. Hence, the true material cost of microchips needs to include the numerous objects that are solely produced and disposed of due to this ultra-cheap internal core that makes those products more marketable/desirable. This is nothing short of an invisible environmental crisis, something made possible by the abundance of cheap digital components whose financial cost bears no relation to its true financial and environmental cost. 

We organised a workshop at ICT4S to explore these costs further.  In this paper we set out the main things we learnt during the workshop, and pose challenges to the LOCO community for addressing this almost invisible, but highly significant, concern.  Not least, we point to the need to educate developers to uncover and exploit the lost potential in the growing mountain of potentially programmable devices all around us, and in our waste streams.

\section{The falling cost of microelectronics}

% Who has done what in this space?

% \subsection{The falling cost of microelectronics} % (fold)
% \label{sub:the_falling_cost_of_microelectronics}

The ICT industry is essential to a vast array of gadgets, products and services that proliferate in numbers beyond most of our imaginations. However, it is the semiconductor microchips, one of the most important components in this digital world, that glues these technologically enabled industries together~\cite{miller2022chip}. The growth of microchips is unprecedented compared to any other industrial sector. Following Moore's law, there has been an efficiency increase  (i.e., transistors per chip) by a million times in 50 years~\cite{mcgregor2022true}. All of the latest AI advances are made possible by the most cutting-edge microchips designed by Nvidia (and others) and fabricated at TSMC, Taiwan. These enormously complex objects consist of over 200 billion transistors (each transistor around 3nm in size!). There is no doubt that the energy and material cost of such an incredible feat of engineering is far beyond anything thus produced to date~\cite{roussilhe2025purer}.  

The `AI hype' has rightly focused our attention on the environmental costs of producing advanced node transistors and also that of their subsequent uses in ever more gigantic data centers. As microchip manufacturers keep pushing further in the name of efficiency (the industry has plans for sub-nm sized transistors in the next 5 years~\cite{ieee-irdsinternational-r}), the true ecological impact behind them will need to be scrutinised even more closely. While future generations of these devices are likely even to need particle accelerators for their production~\cite{boyd2024future}, not all microchips come remotely close to such complexity.  There are, in fact, a vast world of microchips that were originally produced over 30 years ago (many lifetimes in the tech world), but are still in use today. These older generations of microchips (often called mature nodes or `lagging nodes') remain in great demand and are often produced in at least the same volume as they were originally introduced. Incredibly low cost per unit is possible because they are produced in older long established manufacturing facilities (fabs) with little or no further capital investment. %The tools needed to produce them do not require any new RnD and become cheaper to run every passing year.  
In contrast, the advanced nodes are the ones that have entered the market recently and are the product of expensive research and development, requiring state of the are new production facilities. However, it's perhaps inevitable that these expensive advanced nodes of today will become mature nodes in just a few years, and as cost falls, further flood the consumer electronics market. This cycle of lasting production and ongoing impact are often neglected from scrutiny.

% subsection the_falling_cost_of_microelectronics (end)

% Economics

\subsection{Economies of scale} % (fold)
\label{sub:the_economics_of_production}

The longevity of manufacturing old technology ICs is not a random quirk of this industry, but an important feature underpinning its financial viability: since several billion dollars are needed to develop a node generation to maturity, manufacturers keep using them at low cost but large scale for as long as possible. A 350nm microchip, first released in 1995 for example, is not only 100 times less efficient in terms of silicon area and power consumption, but at least 12 generations of node old! With most older generations still actively manufactured, the total number of microchips produced per year keeps accumulating as new generations get added. Historically, a node gets established for commercial production every two years. It takes extensive research and development investment by the microchip industry to actively maintain the development of new nodes at this frequency. The industry trade body has roadmaps for future generations of ever-smaller nodes (sub-nanometer) through to 2037~\cite{ieee-irdsinternational-r}.

\begin{figure}[h]
  \centering
    \includegraphics[width=.9\columnwidth]{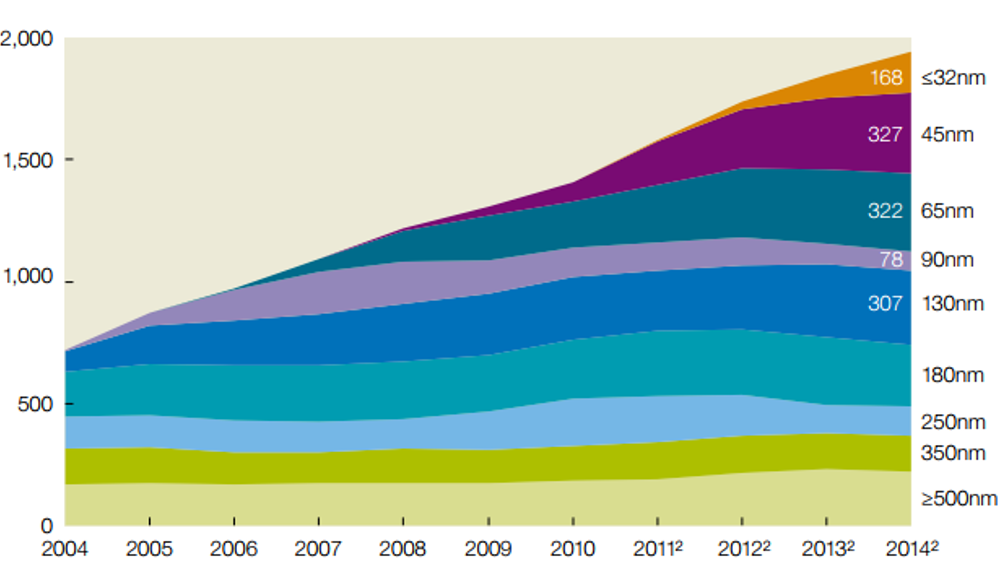}
  \caption{Node production in thousands of wafers per month from 2004 to 2015~\cite{2011mckinsey-on-semi}.}
  \label{fig:figs_kwpm}
\end{figure}

\begin{figure}[h]
  \centering
    \includegraphics[width=.9\columnwidth]{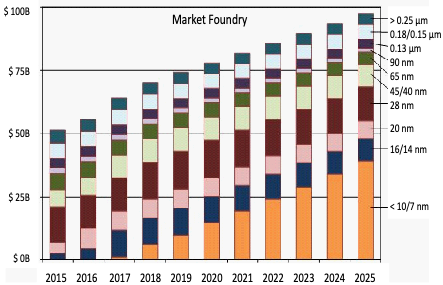}
  \caption{Prediction of revenue share for different nodes show the same trend~\cite{jones2015semiconductor-i}.}
  \label{fig:figs_icrevenue}
\end{figure}

This distinctive effect of microchip production is obvious from available data (Figure~\ref{fig:figs_kwpm}). Note how nodes are established (e.g., 350nm processes), and continue to be manufactured; with each generation layering on the last creating a cumulative growth in the number of devices. Unfortunately, actual data on node-specific production is no longer published by this industry% full of trade secrets%
. However, estimated data shows a similar onward trend (Figure~\ref{fig:figs_icrevenue}); while the bulk of the revenue comes from the more advanced nodes, the mature nodes consistently provides a steady revenue stream~\cite{sachssupply-chain-ga}. 

The economics speak for themselves. The significant up front cost of developing the node is offset by the long lifetime and volume of production and multi-decade sales.  This in part helps explain how it can be possible to produce electronic parts at quantity whose unit cost is in the tens of cents (fractions of a dollar in US pricing), yet may still have computational capabilities in excess of that available to hobbyist programmers in 1980s.

\begin{figure}[h]
  \centering
    \includegraphics[width=.9\columnwidth]{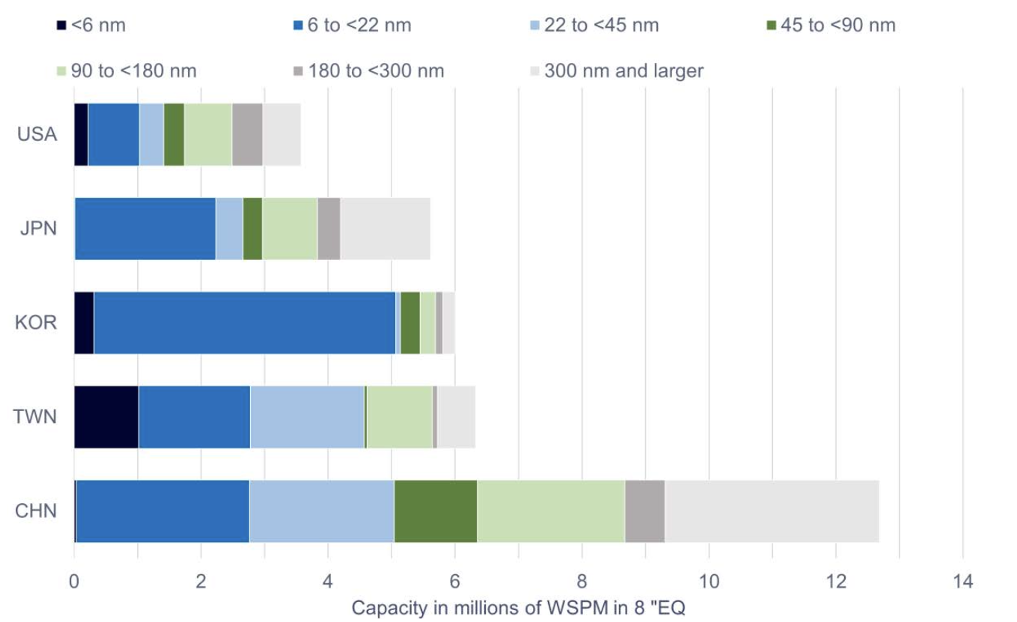}
  \caption{Global distribution of main IC fabs in current production by region, broken down by process~\cite{oecd-th-2025}, 13, Figure~1.}
  \label{fig:figs_geographic}
\end{figure}

Obsolescence and growth are endemic to the financial model.  In terms of economic value, each node generation is either relatively stable or grows in terms of revenue over its lifetime (Figure~\ref{fig:figs_icrevenue}).  This growth despite falling lifetime costs of production for each IC over time, effectively mandating a need for growth in sales volume to maintain profitability.  Industry analysts estimate 500bn ICs were sold in 2025, up from 200bn in 2018.  If we include supporting products, discrete components, passives and sensors, this tops 1.3 trillion units. The bulk of the mature node microchips (>28nm) being produced in China (Figure~\ref{fig:figs_geographic}) confirms the global demand of low-cost electronic goods most often manufactured in the region.  

Market observers suggest that investment in IC foundaries in the last 5--10 years has increased by 280\% driven partly by demand, but also supply chain fears and sovereignty concerns over future supply driven by huge concentrations of manufacturing in specific regions such as Taiwan and China% (Figure~\ref{fig:figs_geographic})
.  According to McKinsey, electronics is the most traded sector of the global economy, representing 4 trillion dollars annually---around 20\% of all global trade---with around 50\% of this going into consumer products, and about \$2.5 trillion as electronic inputs to other sectors~\cite{2011mckinsey-on-semi}.  This abundance of cheap components results in indiscriminate use, hyper-consumption and consequent obsolescence.

% Need something on the cost of a typical electronic gadget

% Capability

% Waste

\subsection{Gold in them thar hills\protect\footnote{Originally used in Mark Twain's novel The American Claimant (1892), this phrase is likely a misquote referring to Gold in the North Georgia Mountains, although far less specific in our context, we use it to represent the materials found in useful concentrations more generally in e-Waste.}} % (fold)
\label{sub:the_practice_of_waste}

Mountains of waste from consumerist societies in the Global North is an enormous and mounting problem.  This has disproportionately high environmental, human and non-human cost is in the low- and middle-income countries. There are numerous examples of electronic waste being shipped to countries where they rot in the environment or are crudely treated with chemicals or are burnt to extract the most valuable and easy to extract components~\cite{clapp2025waste}.  According to UN’s fourth Global E-waste Monitor report (2024) over 62 million tonnes (Mt) of e-waste was produced in 2022, up 82\% from 2010---with projections suggesting this is on track to rise another 32\% to 82Mt by 2030~\cite{un-ewaste-2024}.  Global e-waste monitor observe that only one quarter (22.3\%) of the 2022’s e-waste mass was documented as having been properly collected and recycled. We anticipate a significant lag in this analysis, given the rate of deployment and differing lifetimes of many of these devices%, e.g.\ solar panel end of life is sure to be an emerging issue
. Despite the environmental and humanitarian cost, this represents a huge concentration of underexploited natural resources.

\begin{figure}[h]
  \centering
    \includegraphics[width=.9\columnwidth]{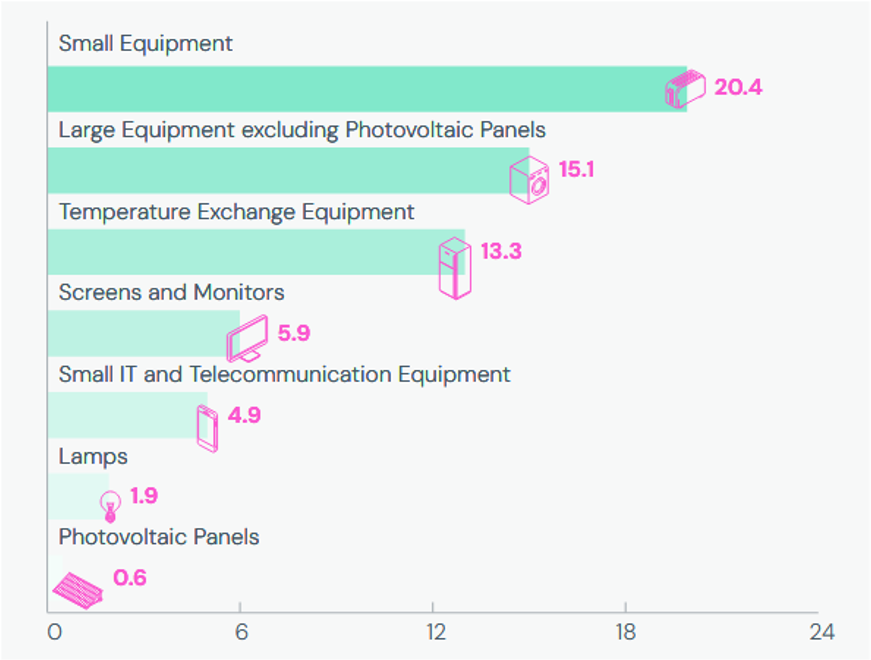}
  \caption {E-waste (in billion Kg) by type of use~\cite{un-ewaste-2024}, 29, Figure~4.   Note the emergence of small equipment (i.e., low-cost electronics: toys, vapes, portable fans) as the largest e-waste sector by weight, surpassing all other categories.}
  \label{fig:figs_ewaste}
\end{figure}

% Urban mining, purity case

\begin{figure}[h]
  \centering
    \includegraphics[width=\columnwidth]{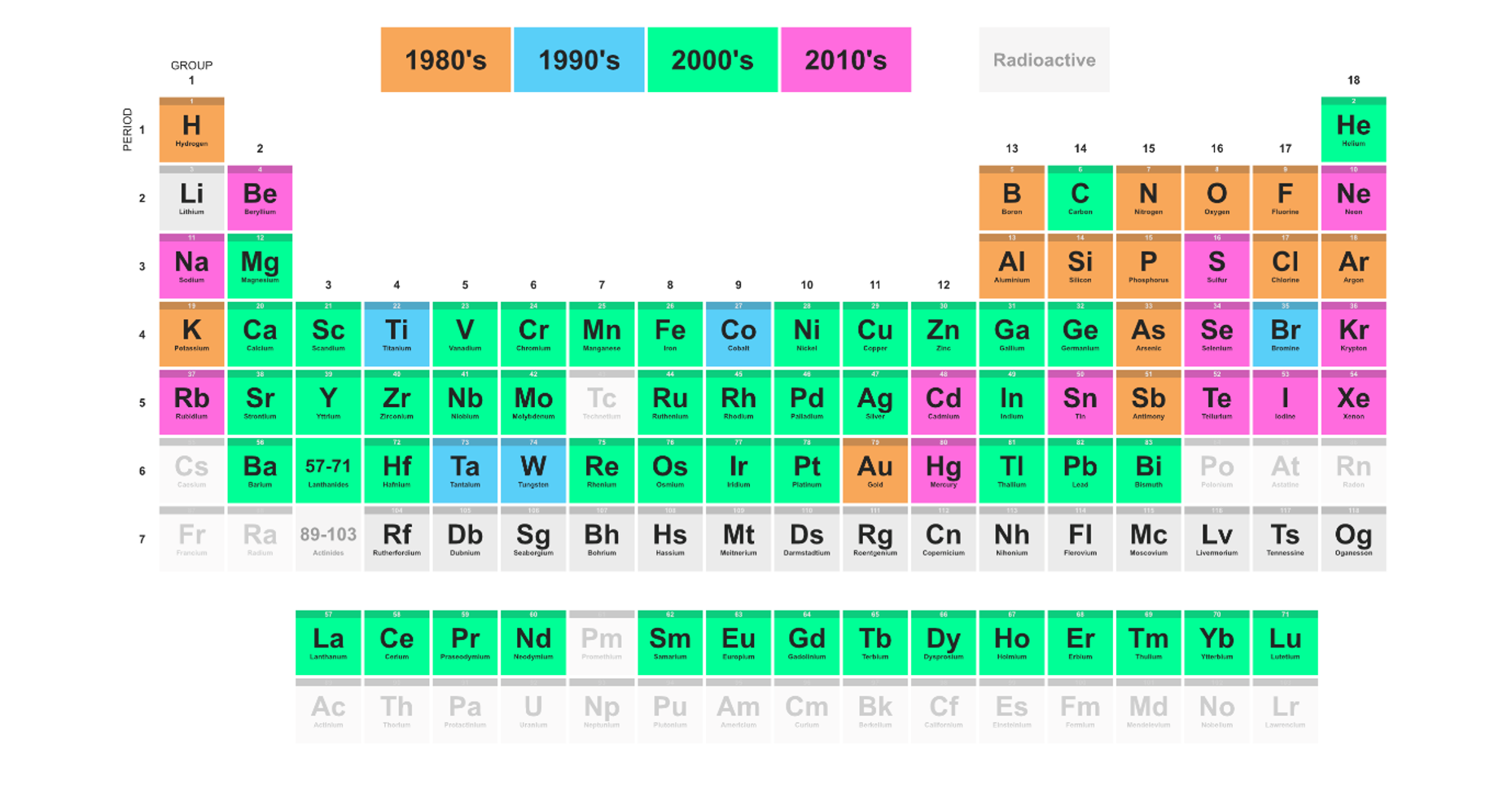}
  \caption{Range of elements needed to make each generation of micro-chip, colour coded by by decade.  Note how the range and complexity of elements has increased over time, with consequent implications for supply chain, embodied emissions, environmental degradation and cost.}
  \label{fig:figs_periodic}
\end{figure}

It's hard to appreciate the extent of the impacts of the electronics industry, without considering the sheer marvel of global supply chains and extractivist practice that collectively enables it.  As generations of microchip require ever finer levels of component integration, and more effective conductors at ever tinier scales on almost atomic levels of precision are needed to link them together, so the industry has come to draw on ever more complex cocktails of materials (Figure~\ref{fig:figs_periodic}).  To such an extent that almost no non-radioactive element hasn't been tried and incorporated.

In recent work, co-authors have written on the escalating levels of purity of material required to enable this industry to continue to progress to finer levels of component integration~\cite{roussilhe2025purer}.  This of course, speaks to remarkable levels of ingenuity, materials science, engineering and enormous quantities of energy, water and material extraction bound up in this upward ratcheting but nonetheless significantly impressive ongoing achievement.  Yet, this also helps explain why rare earth materials are now found at likely higher levels of concentration in legal and illegal electronic waste dumps, than naturally occur in the environment as virgin materials.  At least pragmatically, and not speaking to the difficulty of removing and remanufacturing the very tiny quantities of materials involved, this suggests there is a potential economic case for so called `urban mining' of waste, especially for increasingly pressured and valuable materials such as Copper and Gold.

% Circular economy?

% subsection the_practice_of_waste (end)

\subsection{Design materiality} % (fold)
\label{sub:design_materiality}

\begin{figure}[h]
 \includegraphics[width=.6\columnwidth]{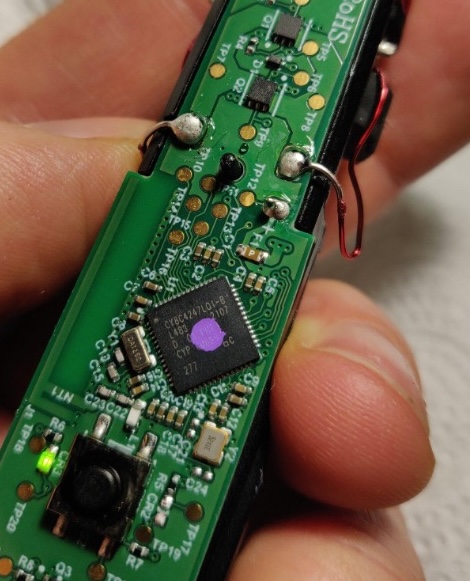}
  \caption{Workshop slide: PCB of an electronic toothbrush.  Next to the push button, a fairly sophisticated microcontroller and supporting interface components.}
  \label{fig:toothbrush}
\end{figure}

Low cost electronics unquestionably enable product designers to create and differentiate their products by adding smart features including blinking lights, displays, rechargable batteries, network connectivity, all at relatively low additional manufacturing cost, especially at volume (Figure~\ref{fig:toothbrush}).

% \begin{figure}[h]
%   \centering
%     \includegraphics[width=.9\columnwidth]{figs/headphone-waste}
%   \caption{Headphones containing tiny specialised audio signal processors, before and `after'.}
%   \label{fig:figs_headphone-waste}
% \end{figure}

% Decided this is off argument (camera ready, AF 14.8)
% \begin{figure}[h]
%   \centering
%     \includegraphics[width=.7\columnwidth]{figs/pointless-screens}
%   \caption{Screens replacing glass refridgerator doors, allowing product marketing, but preventing direct observation.}
%   \label{fig:figs_pointless-screens}
% \end{figure}

% surveillance capitalism?

An important differentiator in connected products is the shift from product revenue derived at the point of sale, to one where the product enables ongoing subscription services from a wider product eco-system; with value derived from long running and likely connected services. The data generated through use of the product also creates additional value; so called `surveillance capitalism'% (Figure~\ref{fig:figs_headphone-waste})
.  This further allows the cost of the product to be offset to gateway into later and ongoing value streams, promoting the integration of smart features that would otherwise be too costly or uncompetitive to integrate.

\begin{figure}[t]
  \centering
    \includegraphics[width=.9\columnwidth]{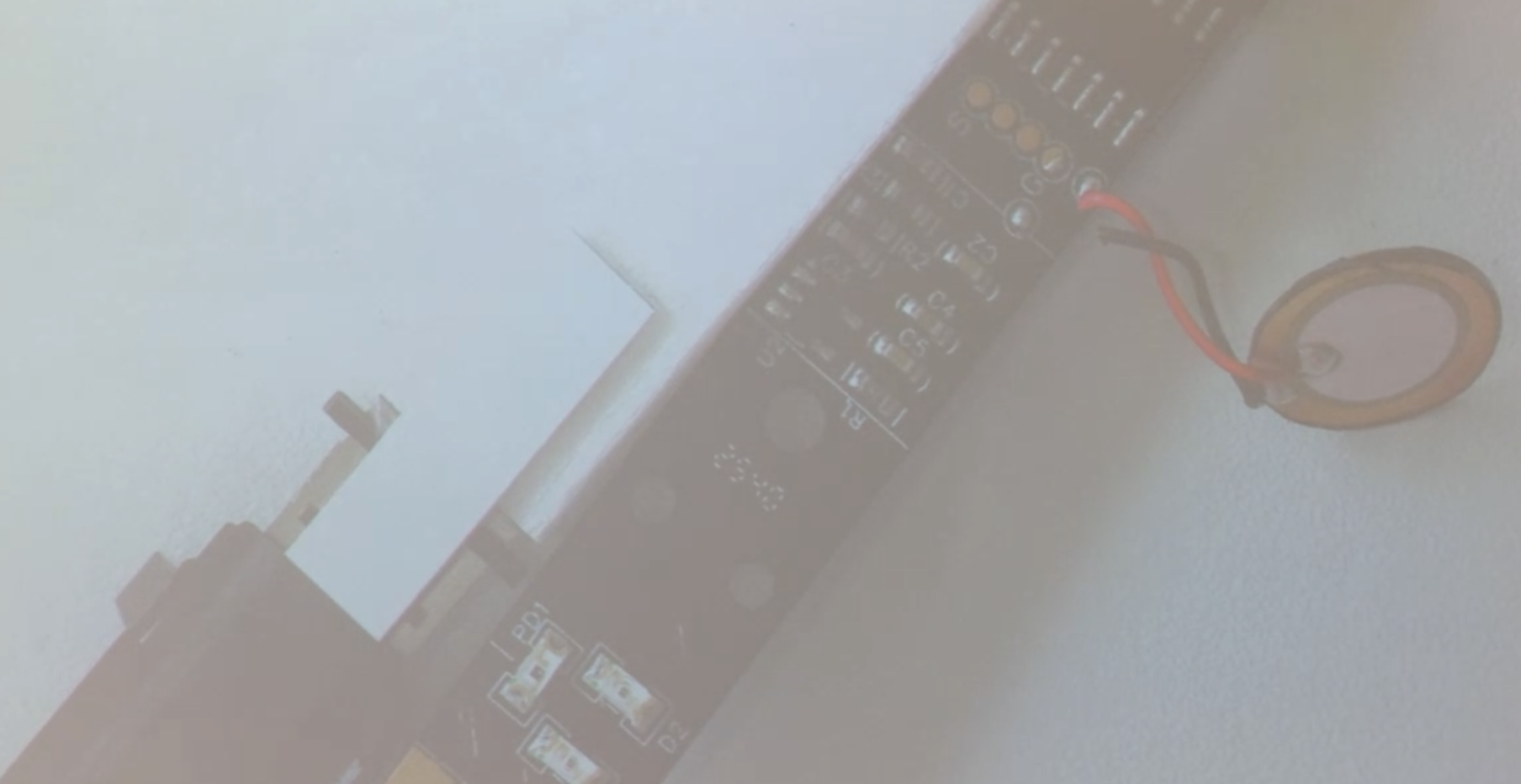}
  \caption{Workshop photo: disassembled digital pregnancy test.  Note how inherently the test is the same chemical process as a conventional test, yet an electronic display replaces the human readable indicator strips by reading the now obscured strips with sensors.  The test \emph{is single use by design}.}
  \label{fig:figs_unboxing-test}
\end{figure}

%In this paper, we were primarily concerned however, with the far more numerous devices that make everyday products `smart', and that are generally speaking, less visible and even disposable.  
Digital is a new design materiality, where aside from traditional concerns in design, the smartness adds a new aesthetic, user experience, or helps differentiate a product or allow it (typically) to be sold at additional cost. 
% focus on design materials/sustainability?
As with fast fashion, low cost clothing that's highly disposable nature, with rapid turnover, excessive waste, and human rights issues in the supply chain which is garnering more and more attention.  We observe what we might term `fast tech', i.e.\ technology so cheap that the products are inherently transient, with short lifetimes or even being  disposable in nature (Figure~\ref{fig:figs_unboxing-test}).

% `Fast-tech'
% military...

It's important to recognise that fast tech is not just a consumer issue, with `things' created for us to buy containing some, many, perhaps an increasing number of electronic components. According to WSTS, consumer products make up around 35\% of mature IC use, but over 70\% of industry, military and government use, due we suspect for the need for established and robust components for many application domains.  %Mature generations of integrated circuits are particularly significant for deployment in military hardware, weapon systems and in space where environmental conditions, low probability of failure and resistance to interference are high priorities.  
It should be obvious that components in military hardware are not ultimately intended to be recycled!

% \begin{figure}[h]
%   \centering
%     \includegraphics[width=.9\columnwidth]{figs/robust-pcb}
%   \caption{Rugged `mature node' integrated circuits.  Shown here, the guidance system for a shoulder launched missile.  It should be obvious that these components are not intended to be recycled.}
%   \label{fig:figs_robust-pcb}
% \end{figure}

% subsection design_materiality (end)

\section{Exploring the lost opportunity} % (fold)
\label{sub:lost_opportunity}

\begin{figure}[h]
    \centering
    \includegraphics[width=.7\columnwidth]{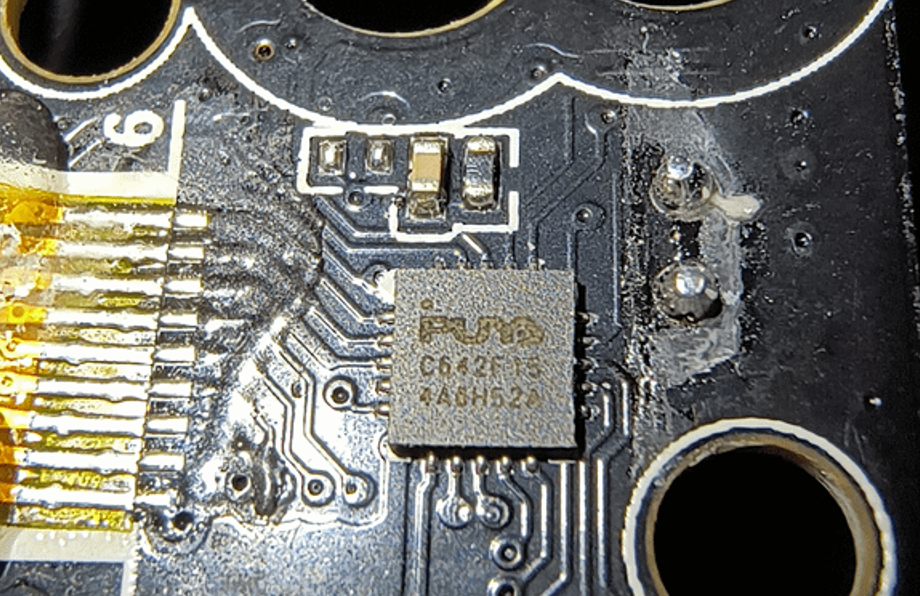}
    \caption{Workshop slide: A `smart vape' PCB, with a 20-pin PUYU micro-controller which controls the lights, heating element, battery and small status LCD display.}
	\label{fig:vapepcb}
\end{figure}

A typical product such as a vape might contain a PCB like that shown in Figure~\ref{fig:vapepcb}.  This contains a 20-pin ARM CORTEX-M0+ micro-controller.  Not lowest specification device in the range.  The cheapest costs around 10\textcent{} in quantity, and features a 24MHz processor with 3Kb of RAM, 10-bit analogue to digital conversion and 18 configurable input-output pins%(see Figure~\ref{fig:vapecircuit}).  
. Informally, it has been demonstrated that a vape can be reprogrammed to host a web server~\cite{bogdan-2025}!

To explore this topic further, in our workshop we disassembled several low cost devices together that we thought could contain embedded electronics: a digital pregnancy test, a non-functioning bike light, a broken hand tool battery charger% which had a charge status light on it
, and some very low cost childrens' toys.  Our goal was to discover what components were integrated, and try to see if we could find where it had originally been made, what capabilities it might have, and how we might think about repurposing it to extend it's useful life.  We shared and discussed what we found, and repeat our key findings here.

% \begin{figure*}[h!]
%     \centering
%     \begin{subfigure}[t]{0.5\textwidth}
%         \centering
%         \includegraphics[height=1.2in]{figs/puya-mcu}
%         \caption{A `vape' PCB.}
%     \end{subfigure}%
%     ~
%     \begin{subfigure}[t]{0.5\textwidth}
%         \centering
%         \includegraphics[height=1.2in]{figs/cortex-m0}
%         \caption{The PUYU MCU is an ARM CORTEX-M0+ micro-controller.}
%     \end{subfigure}
%     \caption{A PCB of a `smart vape' (left).  Note the 20-pin micro-controller (MCU), which controls the lights, heating element, battery and small status LCD display.  Not quite bottom of the range, this MCU still costs only a few 10s of cents.}
% 	\label{fig:vape}
% \end{figure*}

% \begin{figure}[h]{0.5\textwidth}
%     \centering
%     \includegraphics[width=.9\columnwidth]{figs/cortex-m0}
%     \caption{The PUYU MCU is an ARM CORTEX-M0+ micro-controller.}
% \end{figure}
%     \caption{The PUYU MCU schematic.  Not quite bottom of the range, this MCU still costs only a few 10s of cents.}
%     \label{fig:vapecircuit}
% \end{figure*}

\begin{figure}
 \includegraphics[width=.6\columnwidth]{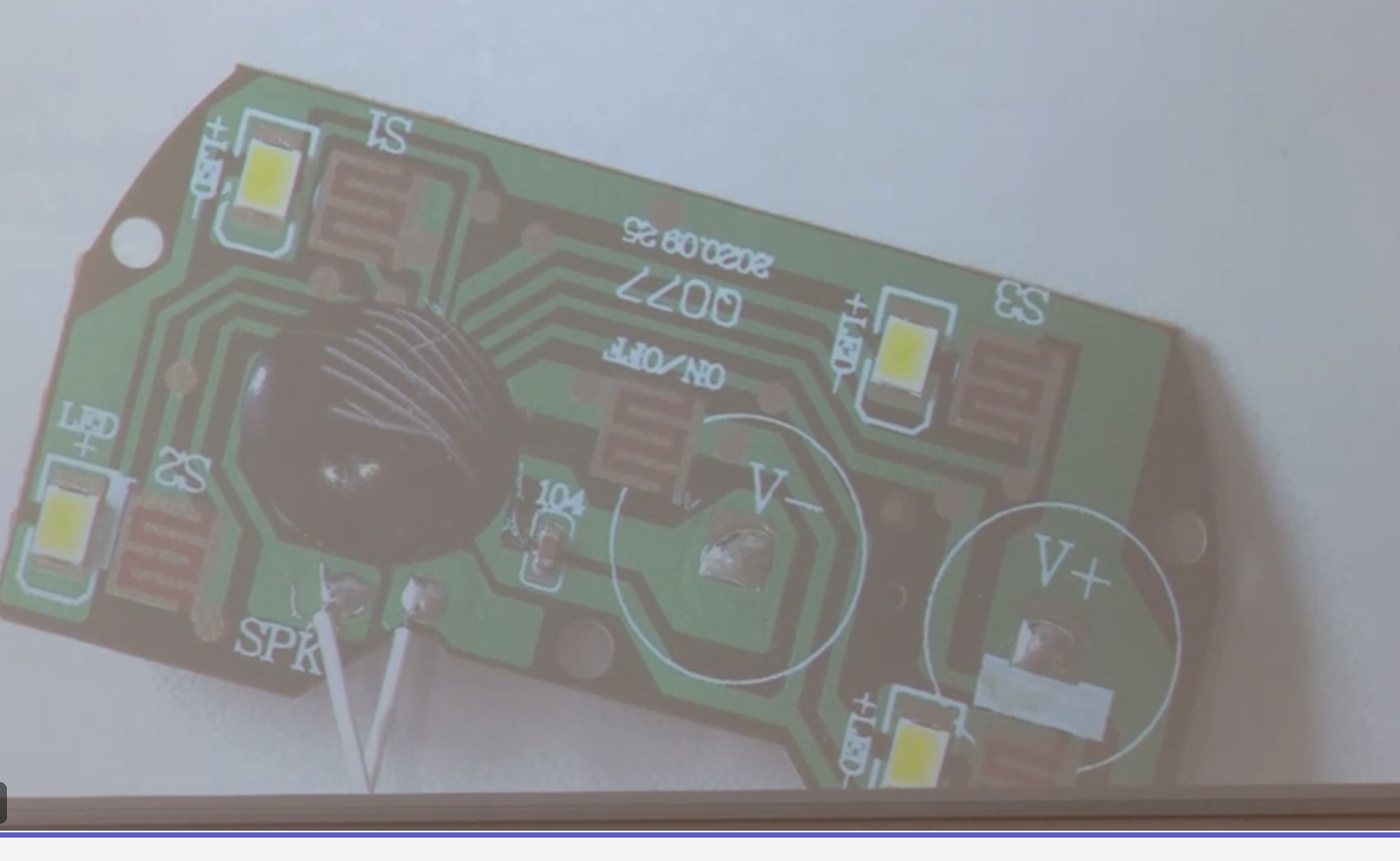}
  \caption{Workshop photo: direct mounted IC in child's toy (under the black resin), so cheap it's surface mounted on the PCB, so really difficult to reuse without destroying it.}
  \label{fig:blob}
\end{figure}

\begin{enumerate}
	\item \textbf{Diversity and quantity of capable ICs.}  We were surprised by the sheer volume of small surface mount IC packages in most of what we disassembled, not just ICs, but many similarly packaged static components.  %Doubtless a result of the production line float solder process even capacitors and resisters were often surface mounted and packaged in small pin packages.  
    %At the `cheap end' of the mature processor scale, w
	We mainly found 16--20 pin micro-controllers---conceptually at least reprogrammable devices if the program code can be overwritten.  Low cost ICs can be application specific non-reprogrammable devices (ASICs) and Field Programmable Gate Arrays (FPGAs), but there is a reasonable argument that, depending on volume and price, it's cheaper to develop using more generic MCUs, than create specific runs of specialist ASICs as the designs, development boards and expertise are more readily available.
	\item \textbf{Really cheap packaging.}  In very cheap production its common to find direct mounted ICs under `resin blobs' (Figure~\ref{fig:blob}).  The resin keeps the circuit safe from interference, air and corrosion, but are clearly intended to cut cost and not designed to be recovered and reused.  We assume it'd be difficult to extract the underlying chip without compromising the packaging; presumably adding something that could be desoldered and reused would increase cost, or bulk, or both.
	\item \textbf{Lack of consistent labelling.} Tracing specific components to its manufacturer took a mixture of web searches for data sheets, and looking up parts lists based on codes printed on IC packaging.  This proved very challenging, not least because any printing on packages of this scale is inherently very small.  Many numbers did not return useful search results, or related to other similarly numbered but irrelevant things.  There's also a surprising number of packages with no markings at all, presumably exactly to avoid the kind of reverse engineering we attempted.
	\item \textbf{Longevity of data sheets.} Due in part to how and where these products were created, we could not always find digital resources relating to the ICs we found; when we could find data sheets, these often related to entire device families, and further investigative work was needed to resolve to a specific device.  Datasheets were mostly in Chinese, but often contained useful circuit schematics that helped us find our specific devices nonetheless.
\end{enumerate}

% subsection lost_opportunity (end)

% Capability
% Waste

% Shenzen/Steve's thing

\section{Lessons learned} % (fold)
\label{sec:lessons_learned}

When we think of low carbon computing, it is often tempting to focus on high end devices, scheduling of large computational jobs, perhaps the lifetime of highly integrated and high end devices such as laptops, servers and smartphones.  Yet, as we hope we've argued, there is a long tail of tiny, almost invisible, computational devices that are being used as a design material that is ephemeral, disposable and cheap.  These exist in absolutely vast quantities.  It is clearly an unbalanced problem: massively easier to manufacture at scale than to recover, reuse, and gain lasting value from.  Once a generation of technology is deployed, it is perhaps destined to become tomorrow's low cost embedded component, made and disposed of for decades thereafter.  Each generation layering and supplementing, but rarely actually replacing, the last.

Borrowing from permacomputing, can we radically rethink computational culture to stem the tide of this wasteful practice?  Quoting from the permacomputing wiki, ``\emph{Caring for the planet also means caring for the material foundations of digital technology: our hardware}'' and drive to ``\emph{produce no waste}''.

As educators of designers, hardware and software engineers, then, can we ensure that we teach skills that are going to make effective use of the hidden gems in the bulging e-Waste scrap heaps?  Certainly there is much to be done in creating products to last, where the digital information to unlock them is maintained long term, i.e., where commons assets like datasheets and developer knowledge remains available long after the devices are manufactured to support their continued use for the decades to come.

How might we make the inherent and true cost more visible as an industry?  We might call for `True cost accounting' to be applied across industry and lobby for new policy.  As with the effective embedding of protections from overuse of hazardous materials has become well regulated and supported (RoHS), so costs beyond the financial (which are let's face it, too cheap) could be surfaced and made more transparent?

More transparency is just the start, Extended Producer Responsibility (EPR) is a mechanism by which producers are held responsible for their products and their packaging beyond the point of sale.  In a drive toward circularity and upcycling of these myriad devices; or even the recovery of the precious materials they contain, clearly more could be done and this might provide a way of cross-subidising the costs of recovery and handling.

Lastly, how do we drive for overall sufficiency in microchip production.  It is clear our society is producing massively more than we need, or to sustain essential services of life.  Pushing for sufficiency will require thinking much more deeply about how to address the underlying economics, political and capital drivers that atomise precious materials future generations will be glad of.

% section lessons_learned (end)

% permacomputing

% better practice

% skilling up

\bibliographystyle{ACM-Reference-Format}
\bibliography{abundance.bib}

@book{un-ewaste-2024,
	author = {{UN}},
	institution = {{UN}},
	month = {November},
	publisher = {{UNINTAR}},
	title = {The global E-waste Monitor 2024 -- Electronic Waste Rising Five Times Faster than Documented E-waste Recycling},
	url = {https://ewastemonitor.info/the-global-e-waste-monitor-2024/},
	volume = {2},
	year = {2024}}

@book{clapp2025waste,
	address = {John Murray},
	author = {Clapp, Alexander},
	isbn = {1399803123},
	month = {February},
	publisher = {Hachette UK},
	title = {Waste wars: the wild afterlife of your trash},
	year = {2026}}

@article{oecd-th-2025,
	author = {{OECD}},
	journal = {Oecd Science, Technology and Industry Policy Papers},
	month = {December},
	number = {188},
	title = {The Chip Landscape: Geographical Distribution of Wafer Fabrication Capacity},
	year = {2025}}

@electronic{sachssupply-chain-ga,
	author = {Jason Sachs},
	title = {Supply Chain Games: What Have We Learned From the Great Semiconductor Shortage of 2021? (Part 3)},
	url = {https://www.embeddedrelated.com/showarticle/1489.php},
	urldate = {10 December 2022}}

@techreport{jones2015semiconductor-i,
	address = {semi.org},
	author = {Handel Jones},
	institution = {International Business Strategies (IBS)},
	month = {August},
	title = {Semiconductor Industry from 2015 to 2025},
	url = {https://www.semi.org/en/semiconductor-industry-2015-2025},
	year = {2015}}

@techreport{2011mckinsey-on-semi,
	author = {{McKinsey and Company}},
	institution = {{McKinsey and Company}},
	month = {Autumn},
	number = {1},
	title = {McKinsey on Semiconductors},
	url = {https://www.mckinsey.com/~/media/mckinsey/dotcom/client_service/semiconductors/pdfs/mosc_1_revised.ashx},
	year = {2011}}

@article{banfield-nwachi2025its-cheap,
	author = {Mabel Banfield-Nwachi},
	journal = {The Guardian},
	month = {June},
	title = {It's cheap but it's not disposable': why fast tech is a growing waste problem},
	url = {https://www.theguardian.com/technology/2025/jun/24/its-cheap-but-its-not-disposable-why-fast-tech-is-a-growing-waste-problem},
	year = {2025}}

@electronic{StatistaMarketInsights,
	author = {{Statista Market Insights}},
	title = {Semiconductors - Worldwide by Volume},
	url = {https://www.statista.com/outlook/tmo/semiconductors/worldwide#volume},
	urldate = {Jul 2025},
	year = {2025}}

@electronic{bogdan-2025,
	author = {Bogdan Ionescu},
	lastchecked = {14 August 2026},
	title = {Hosting a Website on a Disposable Vape},
	url = {https://bogdanthegeek.github.io/blog/projects/vapeserver/},
	urldate = {13 September 2025},
	year = {2025}}

@article{boyd2024future,
	author = {Boyd, John},
	journal = {IEEE Spectrum},
	number = {7},
	pages = {28--33},
	publisher = {IEEE},
	title = {Is the future of Moore's law in a particle accelerator?: Wiggling electrons could turbocharge EUV lithography},
	volume = {61},
	year = {2024}}

@webpage{ieee-irdsinternational-r,
	author = {{IEEE IRDS}},
	lastchecked = {24 June 2026},
	title = {International Roadmap for Devices and Systems\textsuperscript{TM}},
	url = {https://irds.ieee.org},
	urldate = {2023},
	year = {2023}}

@article{roussilhe2025purer,
	author = {Roussilhe, Gauthier and Pirson, Thibault and Bol, David and Mitra, Srinjoy},
	journal = {arXiv preprint arXiv:2509.18768},
	title = {{\bf In press}. Purer than pure: how purity reshapes the upstream materiality of the semiconductor industry},
	year = {2025}}

@article{mcgregor2022true,
	author = {McGregor, Jim},
	journal = {Forbes},
	month = {October},
	title = {The True Nature Of Moore's Law---Driving Innovation For The Next 50 Years},
	url = {https://www.forbes.com/sites/tiriasresearch/2022/10/07/the-true-nature-of-moores-law--driving-innovation-for-the-next-50-years/},
	year = {2022}}

@book{miller2022chip,
	author = {Miller, Chris},
	publisher = {Simon and Schuster},
	title = {Chip war: The fight for the world's most critical technology},
	year = {2022}}

\end{document}